\documentclass{article}
\usepackage[utf8]{inputenc}
\usepackage[margin=1.4in]{geometry}
\usepackage{graphicx}
\usepackage{amsmath, amsfonts}
\usepackage{setspace}
\usepackage{multirow}
\usepackage{graphicx, epstopdf}
\usepackage{mathrsfs}
\usepackage{mdwlist}
\usepackage{caption}
\usepackage{rotating}
\usepackage{footnote}
\usepackage{lscape}
\usepackage[table,xcdraw]{xcolor}
\newcommand{\test}[1]{\expandafter\hat#1} 
\usepackage{algorithm}
\usepackage{algpseudocode}
\usepackage{bm}
\usepackage{rotating}
\usepackage{pdflscape}
\usepackage{subfig}
\usepackage[numbers]{natbib}
\usepackage[countmax]{subfloat}

\title{Computationally efficient surrogate-based optimization of coastal storm waves heights and run-ups}
\author{Theodoros Mathikolonis \footnote{Department of Statistical Science, University College London, UK (theodoros.mathikolonis.13@ucl.ac.uk)} \,\,\,\ Volker Roeber \footnote{E2S UPPA Chair HPC-Waves, Laboratoire SIAME, Univ Pau \& Pays Adour, Anglet, France (volker.roeber@univ-pau.fr)} \,\,\,\ Serge Guillas \footnote{Department of Statistical Science, University College London, UK (theodoros.mathikolonis.13@ucl.ac.uk)}}

\date{October 2019}

\begin{document}

\maketitle
\begin{abstract}
The computational burden of running a complex computer model can make optimization impractical. Gaussian Processes (GPs) are statistical surrogates (also known as emulators) that alleviate this issue since they cheaply replace the computer model. As a result, the exploration vs. exploitation trade-off strategy can be accelerated by building a GP surrogate. In this paper, we propose a new surrogate-based optimization scheme that minimizes the number of evaluations of the computationally expensive function. Taking advantage of parallelism of the evaluation of the unknown function, the uncertain regions are explored simultaneously, and a batch of input points is chosen using Mutual Information for Computer Experiments (MICE), a sequential design algorithm which maximize the information theoretic Mutual Information over the input space. The computational efficiency of interweaving the optimization scheme with MICE (optim-MICE) is examined and demonstrated on test functions. Optim-MICE is compared with state-of-the-art heuristics such as Efficient Global
Optimization (EGO) and GP-Upper Confidence Bound (GP-UCB). We demonstrate that optim-MICE outperforms these schemes on a large range of computational experiments.
\end{abstract}

\section{Introduction}
Storm surges cause coastal inundation due to setup of the water surface resulting from atmospheric pressure, surface winds and breaking waves. They occur during tropical cyclones, also known as typhoons or hurricanes, and are referred as the major geophysical risks associated with coastal areas due to the large numbers of casualties and damage \cite{ellis2015perspectives,feng2018storm}.\\
\\
Typhoons are subject to climate change-related influences, such as warmer sea surface temperatures and sea level rise \cite{bengtsson2009will, lin2012physically, wang2015quantifying}. Despite the fact that modelling storm surge events under different climate scenarios is not straightforward, changes in the storm surges' intensities and frequencies in a warmer climate have been investigated with high-resolution climate models \cite{bengtsson2009will, lin2012physically}. However, different studies have shown that, statistically, weather-climatic extremes (e.g. wind, precipitation) are still sometimes over-or under-estimated due to the difficulty to evaluate them in climate simulations, adding another layer of uncertainties to assess future scenarios \cite{muis2016global, romero2017climate}. Therefore, it is crucial to quantify carefully the local impact of storm waves following these regional climate forcings in order to estimate future (and possibly increasing) storm wave risks.\\
\\
About half of the strongest typhoons in the Western North Pacific hit the Philippines as the country lies in the most tropical cyclone-prone waters on Earth \cite{ribera2008historical,takagi2017track}. On 8 November 2013, one of the strongest tropical storm events ever recorded in the Philippines, Typhoon Haiyan, damaged billions worth of agriculture and infrastructure and, caused 6340 casualties. The town of Hernani, which was completely destroyed, got the attention of multiple research groups \cite{roeber2015destructive, may2015block, soria2018surf}. What is interesting about the situation of Hernani, is that the broad fringing coral reef near its coast was expected to protect the coastal communities \cite{ferrario2014effectiveness} however, in the case of Hernani, the reef exacerbated the damage from energetic waves. Furthermore, as explained by Roeber \& Bricker \cite{roeber2015destructive}, the nature of this event has remained a mystery for some time. Initially, it was assumed that the damage was caused by a storm surge, but the water level due to barometric-, wind-, and wave-induced set-up would not have carried the destructive power - especially because storm surges move rather slow. The possibility of a meteo-tsunami or a near-field tsunami was also precluded as the propagation speed of the storm system was not in phase with the local wave celerity and no tsunami-generating earthquake or landslide activity was reported.\\
\\
 Various studies state that the strong wind from Typhoon Haiyan pushed enough water onshore causing a storm surge height of 6m \cite{mori2014local}. The areas most affected by the storm surge were close to Tacloban, where the bathymetric conditions favored the generation of such a massive storm surge. Surprisingly, destructive long-period bores struck the town of Hernani. Though the actual storm surge was rather small, wave run-up heights of over 7m have been reported around Hernani's Pacific coastline \cite{soria2016repeat}. The duration of the flooding caused by each bore was on the order of a minute and therefore much longer than the period of regular storm waves.  The tsunami-like waves, which resulted from the wave-breaking process over the steep reef face, have been shown to be generated by the energetic surf beat under some particular conditions that favored the generation of infra-gravity waves (e.g. reef geometry, wave patters whose ranges of variations will depend upon climate forcing). Due to the fact that a similar tsunami-like flood was only reported during the 12 October 1897 typhoon, coastal risk assessment and evacuation plans did not consider the existence of this rare but disastrous event. The phenomenon in Hernani was successfully captured by \cite{roeber2015destructive} using a Boussinesq-type phase-resolving wave model, called BOSZ, able to reproduce the transfer of short wave energy into energetic long-period infra-gravity waves \cite{roeber2012boussinesq}. \\
\\
In general, the impact of inundation and coastal flooding events is often quantitatively assessed by the maximum wave run-up. The wave run-up is defined as the maximum vertical extent of a wave's up-rush on a beach or structure above a known reference level such as a chart datum or simply the mean water level. Understanding the conditions that caused the wave run-up and obtaining accurate predictions are of utter importance for coastal flooding hazard predictions. Due to that, wave run-up has been extensively studied either to observe the interaction between tsunami-like waves and fringing reefs \cite{yao2018study, ning2019study}, or to understand the influence of infragravity (long) waves \cite{shimozono2015combined, montoya2018tsunami}, and the effects of climate change on coastal flooding risk \cite{lin2012physically, quataert2015influence}.\\
\\
Various statistical approaches are also widely used to study the wave run-up. Extreme Value Theory (EVT) is applied to analyse the storm peak significant wave height and estimate the failure probabilities of offshore structures \cite{northrop2017cross}. It is also used to evaluate the risk of storm tides considering projected changes to cyclone behaviour and the impact of wave setup (an effect of breaking waves at the coast) \cite{mcinnes2003impact}. Hakkou \textit{et al.} \cite{hakkou2019assess} estimated the extreme total water level using classical EVT theory taking into account the differences of coastal morphology in order to assess the flooding. Using a point-wise and spatial statistical model to build a high-resolution hindcast data-set, Sartini \textit{et al.} \cite{sartini2017spatio} investigated the spatial variability of extreme significant wave heights aiming to improve the understanding of the processes governing wave climate. The combined impact of wave height and other variables (water level, wind waves etc.) has been studied by applying advanced probabilistic approaches \cite{jonathan2014non, leijala2018combining}. Multivariate copula functions are also widely used to model the dependence between wave height, wave period, water level and storm duration and to analyse their extremes \cite{de2005modelling, li2014probabilistic, rueda2016multivariate}. The above studies are based on data-driven methods. \\
\\
To overcome the lack of data and explore various scenarios, numerical models are widely used in this field. They provide valuable information about the coastal flooding events as they can compute the relevant physical processes efficiently and capture the extreme environmental events successfully. But because they are computationally demanding, further studies and more detailed analysis, such as sensitivity analysis, become impractical \cite{sarri2012statistical}. To overcome this issue, empirical equations are used as a simpler method to estimate run-up elevation during extreme surge events. Due to the fact that these methods are mostly derived from specific data-sets and represent a range of commonly encountered conditions, their applicability to more complex geophysical situations is limited and within their calculations can possibly include a substantial error \cite{park2016empirical, ji2018empirical}.\\
\\
Recent studies highlight the need for further understanding the risk associated with natural hazards in order to improve coastal resilience \cite{behrens2015new}. The advancement in computational methods has given the opportunity to reduce the computational burden of numerical simulations and perform challenging studies such as the quantification of the uncertainty in the predictions of hazard characteristics. Statistical emulators, which have been shown to be a prominent solution and alleviate the computational cost issues, are starting to be used in this field. Sarri \textit{et al.} \cite{sarri2012statistical} used a Gaussian process statistical emulator to accurately approximate the landslide-generated tsunami model built by Sammarco \& Renzi \cite{sammarco2008landslide} whereas Beck \& Guillas \cite{beck2016sequential}, proposed an improved experimental design to efficiently build an emulator for a tsunami model. Statistical emulators have been also used in more recently to quantify tsunami hazard over the North Atlantic Ocean \cite{salmanidou2017statistical} and Cascadia \cite{guillas2018functional}, to explore how eruption source parameters affect volcanic radiative forcing \cite{marshall2018exploring}, to quantify the uncertainty in Manning's friction parametrization applied to predict the sea surface elevations for the 2011 Japanese tsunami \cite{sraj2014uncertainty}, to identify different sources of uncertainty contributed in volcanic ash transport and dispersion simulators \cite{spiller2014automating}, as well as to perform fast and efficient predictions of estuarine hydrodynamic variables, such as the water levels and non-tidal residual, in the USA Pacific Northwest \cite{parker2019emulation}. \\
\\
A statistical emulator is often seen as an integral part in an optimization algorithm, which aims to maximize a numerical simulator, especially in situations where the traditional mathematical approach can not be taken due to the complexity of the simulator. Stefanakis \textit{et al.} \cite{stefanakis2014can} used an active experimental design strategy to find the combination of parameters which will give the maximum run-up amplification of typical tsunami waves with the lowest possible model runs. The method explores the entire input space and focuses on the uncertain region, which contains the targeted location of the maximum, using the Active Learning MacKay (ALM), an active experimental design. At each time step, each combination of the input parameters is examined considering its predictive variance (a measure of uncertainty of the point predicted). The same optimization technique was integrated into a methodology of predicting and optimizing the layouts of wave energy converters in a wave farm \cite{sarkar2016prediction}.\\
\\
This study focuses on finding the worst case scenario from extreme coastal storm and infra-gravity waves - such as the ones that destroyed the town of Hernani - with a minimum computational effort. A suite of storm waves that lead to extreme wave run-up and bore heights are studied to understand sensitivities to inputs and to identify the conditions that will create possibly large storm wave run-ups at the local level. To do that, optim-MICE, a newly developed optimization scheme, which is based on the Gaussian Process and an information theoretic mutual information measure, is used to find the combination of reef and wave spectral parameters that affect surf beat run-up and forces on coastal structures \cite{mathikolonis2019surrogate}. Taking advantage of the computational efficiency of the optim-MICE algorithm, the maximum run-up and bore height are estimated using lowest possible number of evaluations and thus keeping the computational requirements to the minimum. This is the first maximization of storm surge heights and run-ups using surrogate-based optimization.\\   
\\
The paper is organized as follow. Section 2 describes the two idealised set-ups. Section 3 gives the background about Gaussian Process emulation and experimental design. An explanation of the optim-MICE algorithm is given in Section 4. The computational efficiency of the optim-MICE algorithm compared with state-of-the-art heuristics is presented in Section 5. The optimization results and a local sensitivity analysis around the maximum bore height and run-up are performed in Section 6. We finally conclude in Section 7. 

\maketitle
\section{Storm Surge Simulations}
\begin{figure}[ht]
\minipage{\textwidth}
\centering
  \includegraphics[scale=0.25]{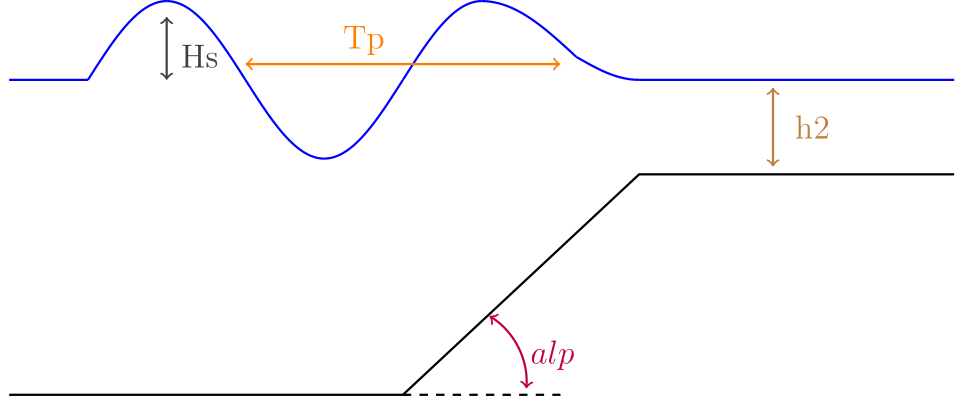}
    \vspace*{0.10cm}
\endminipage\hfill
\captionof{figure}{Bore height: Bathemetric profile of the experimental set-up.}
\label{fig:geo1} 
\end{figure}

\begin{figure}[ht]
\minipage{\textwidth}
\centering
  \includegraphics[scale=0.25]{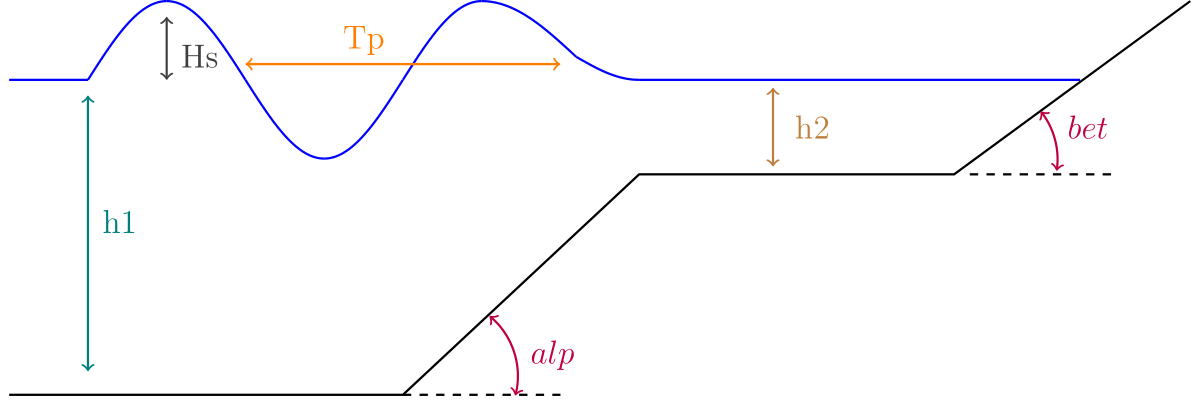}
    \vspace*{0.10cm}
\endminipage\hfill
\captionof{figure}{Wave Run-up: Bathemetric profile of the experimental set-up.}
\label{fig:geo2} 
\end{figure}

The numerical simulations are performed using BOSZ (the Boussinesq Ocean and Surf Zone model), a Boussinesq-type phase-resolving wave model, which can handle the generation of large nearshore waves and their breaking process over an abrupt fringing reef \cite{roeber2012boussinesq}. BOSZ has successfully captured the Typhoon Haiyan storm surge event, which happened back in Philippine Islands as it is able to compute the tsunami-like surf beat over the reef near Hernani. The massive bore that destroyed Hernani was an extreme infra-gravity surge, which resulted from the transfer of energy from short swell waves to long infra-gravity waves initiated by wave breaking at the reef edge. The capability of the BOSZ model in handling these wave processes provides the baseline for studying different aspects of storm wave extrema such as the maximum bore height and the maximum run-up. The maximum bore height is defined as the highest free surface elevation over the section of the horizontally flat reef, whereas the maximum run-up is the highest vertical elevation on the beach, to which the waves reach. Both measures are governed by different processes. To study both extremes the model input needs to be optimized separately. \\
\\
In practice, one of the challenges in an optimization task is to decide when to stop the iterative strategy. A number of different empirical stopping criteria have been proposed in the literature where the entire optimization procedure is stopped once a rule is satisfied \cite{azimi2012hybrid, stefanakis2014can}. But in reality, the computational resources are limited and only a certain number of function evaluations can be performed. Due to the computational complexity of BOSZ, we fix the total number of function evaluations, and therefore the computational time, able to perform in both tasks. The predefined limit is in line with the algorithm settings chosen in \cite{mathikolonis2019surrogate} ensuring the computational efficiency of optim-MICE regardless dimensionality and function complexity. \\
\\
The two main inputs of BOSZ are the bathymetry/topography and the offshore wave spectrum. For each extreme measure a different simplified bathymetric profile is created whereas the wave generation for the wave input is based on the empirical JONSWAP distribution \cite{hasselmann1973measurements}, which depends solely on the values of significant wave height, $Hs$, and peak period, $Tp$. Tables \ref{tab:1} and \ref{tab:2} show the physical parameters incorporated in each optimization task. The range of the parameters was selected to account for reasonable and physically possible combinations. It was not the goal to utilize the maximum values of the parameters. Fig. \ref{fig:geo1} and \ref{fig:geo2} show the simplified bathymetric profiles used in the optimization task. 

\begin{table}[ht]
\centering
\caption{Physical Parameters for Bore Height}%%%Table caption goes here
\label{tab:1}
\begin{tabular}{lll}%%%The number of columns has to be defined here
\hline
Parameter & Measure & Range  \\
\hline
alp: Fore-reef slope  & degrees & 3 - 33.33   \\
h2: Water depth over the reef & m & 0.01 - 3  \\
Hs: Significant wave height & m & 1 - 5\\
Tp: Peak Period & s & 10 - 18 \\\hline
\end{tabular}
\vspace*{-4pt}
\end{table}%%%End of the table

\begin{table}[ht]
\centering
\caption{Physical Parameters for Run-Up}%%%Table caption goes here
\label{tab:2}
\begin{tabular}{lll}%%%The number of columns has to be defined here
\hline
Parameter & Measure & Range  \\
\hline
alp: Fore-reef slope  & degrees & 3 - 33.33   \\
h1: Water depth offshore & m & 50 - 75 \\
h2: Water depth over the reef & m & 0.01 - 3  \\
bet: Beach slope & degrees & 1 - 30 \\
Hs: Significant wave height & m & 1 - 5\\
Tp: Peak Period & s & 10 - 18 \\\hline
\end{tabular}
\vspace*{-4pt}
\end{table}%%%End of the table

\maketitle
\section{Gaussian Process Bandit Setting}
The current study addresses the problem of sequentially optimizing an unknown (black-box) function, here the BOSZ computer model for either the Bore Height or the Run-Up. Let $f :\mathcal{X} \to \mathbb{R}$ be our unknown function with $\mathcal{X} \subseteq \mathbb{R}^d$, compact and convex. The aim is to find with the lowest possible number of function evaluations, the maximum of the unknown function
\begin{equation}
f(x^*) = \underset{x \,\in \,\mathcal{X}}{\max} \,\,  f(x),
\end{equation}\
where $x^*$ denotes the true optimum of $f$. At each iteration $t$, a batch of $K$ input points ($x_t^k$) in $\mathcal{X}$ are chosen and then, the function values at these locations are simultaneously obtained. A common sequential design strategy to optimize a black-box function is the Bayesian Optimization. Although the Bayesian philosophy is not fully adopted in this study, the Bayesian Optimization framework is followed to tackle the sequential batch optimization problem: a probabilistic model is firstly built and then, an acquisition function is used to determine the new input points by satisfying some optimality criterion.\\ 
\\
A Gaussian Process (GP) regression, which is the most well known probabilistic model used in black-box optimization due to its flexibility and tractability \cite{jones1998efficient}, is used as a statistical surrogate model, also known as emulator or meta-model. Specifically, a computer model is often a complex and a computational expensive black-box function and that makes activities such as optimization tasks and sensitivity analysis impractical. To overcome this issue, a statistical surrogate model is built to approximate the input-output behaviour and accurately represent the analytical model, which is often a complicated function form \cite{ohagan2006}.
 
\subsection{Gaussian Process}
A GP enforces implicit properties of the unknown function $f$ without relying on any parametric assumptions. By modelling $f$ as a sample from a GP, a certain level of smoothness, and correlation between nearby locations, can be formalized. A GP is a continuous extension of multidimensional normal distribution, defined by the mean function $m(x)$ and the covariance function $c(x,x')$. The unknown function $f$ can be thought as a Gaussian random function and its output $y=f(x)$ can be described as GP:
\begin{align}
f(x) &\sim GP(m(x), c(x,x')),\\
\text{where}\,\,\,\,\ m(x) &= \mathbb{E}[f(x)],\\
\text{and}\,\,\,\,\ c(x,x') &= \mathbb{E}\left[\left(f(x) - m(x)\right)\left(f(x') - m(x')\right)\right].
\end{align}\ 
The covariance function is defined as $c(x,x') = \sigma^2K(x,x')$: a product of the process variance ($\sigma^2 > 0$) and a covariance matrix $K$. Unlike the mean function which can be chosen freely, an arbitrary covariance function, in general, is not valid as it has to be a symmetric positive semi-definite function. The covariance function - also known as kernel - is the heart of the GP as it encodes the properties of $f$. A range of the different class of covariance functions can be found in Rasmussen \textit{et al.} \cite{rasmussen2005} but the most commonly-used are:   
\begin{itemize}
\item \textit{Separable Power Exponential} with $\xi = (l_1, \dots, l_d)^T \in \mathbb{R}^{d}_{+}$ where the length-scale for the \textit{i}th input dimension is $l_i >0$. The degree of smoothness is controlled by $0 < p \leq 2$ with a typical default choice 2 \cite{santner2003,gramacy2012}, 
\begin{equation}
K(x,x'|\xi)) =  \prod_{i=1}^{d} \exp \left\{ - \frac{\|x-x'\|^{p}}{l_i} \right\},
\end{equation}
\item \textit{Mat\'ern} with length-scale $l_i>0$, $\xi = (l_1, \dots, l_d)^T$, and parameter $\nu$ which controls the smoothness level,
\begin{equation}
K^*(x,x'|\xi,\nu) = \prod_{i=1}^{d}\frac{1}{2^{\nu-1}\Gamma(\nu)} \left(\frac{2\nu^{1/2} \| x - x'\|}{l_i}\right)^\nu J_\nu \left(\frac{2\nu^{1/2} \|x - x'\|}{l_i} \right),  
\end{equation} 
where $\Gamma_\nu$ is the Gamma function for $\nu$ and $J_{\nu}$ is a modified Bessel function of order $\nu > 0$. A common choice for the degree of smoothness are $\nu = 3/2$ and $\nu = 5/2$ \cite{stein1999,minasny2005,rougier2009}.
\end{itemize}\
The uncertain parameters can be modelled using a Bayesian approach \cite{gramacy2009,handcock1993,rasmussen2005}. However, the current study adopts the Design and Analysis of Computer Experiments (DACE) framework proposed by Sacks \textit{et al.} \cite{sacks1989}. The GP regression not only offers a prediction at a new input point but also provides an estimate of the uncertainty in that prediction \cite{sacks1989}. Conditionally on the training outputs after $T$ iterations,  $Y_T = \left[y_t, \dots, y_T \right]^T$ at points $X_T=\left\{x_t, \dots, x_T \right\}$, the process is still a GP and the predictive distribution of the output at a new input points $x$, also known as test points, is a multivariate normal with mean $\test{y}(x) $ and variance $\test{s}^2(x)$: 
\begin{align}
\test{y}_{T+1}(x) &= k_T(x)^T(K_T + \sigma^2I)^{-1}Y_T\\
\text{and} \,\ \test{s}_{T+1}^2(x) &= k(x, x)-k_T(x)^T(K_T+\sigma^2I)^{-1}k_T(x)
\end{align}  \
where $k_T(x) = \left[k(x_1, x), \dots, k(x_T,x)\right]^T$ is the vector of covariances between the input points already chosen and $x$ and $K_T=\left[(x,x')\right]_{x,x' \in X_T}$ is the covariance matrix. These equations 
are also known as kriging equations.

\subsection{Experimental Design}
The computational complexity of a computer model often only allows to perform a limited number of runs. To minimize the computational cost, and maximize the information gained about the unknown function, an experimental design is used where an efficient set of input points ($n$ design points) is chosen strategically based on various optimization techniques \cite{johnson1990,sacks1989,santner2003}. The experimental design strategies are classified mainly into two categories: space-filling designs and adaptive designs.\\
\\
Space-filling designs choose all the design points before computing any function evaluation. All regions of the design space are treated as equally important and as a result, a certain amount of computational time is wasted because unnecessary regions are explored. Examples of space-filling designs are uniform designs, maximin and minimax distance design and Latin Hypercube designs (LHD) \cite{sacks1989, simpson2001}. The adaptive designs, compared with space-filling designs, can be sometimes computationally expensive but often more effective \cite{beck2016sequential}. In adaptive designs, only the most informative input points are included in the training data set by optimizing, at each step of the experimental design process, a specific design criterion \cite{beck2016sequential,gramacy2009,santner2003}. The design points are chosen sequentially, often one-at-a time or in batches, from regions where uncertainty is large. The adaptive designs measure an information gain quantity which can serve as a criterion for designing computer experiments \cite{currin1988}. Examples of adaptive designs are the Active Learning MacKay (ALM) \cite{mackay1992}, the Active Learning Cohn (ALC) \cite{cohn1996} and the Mutual Information for Computer Experiments (MICE) \cite{beck2016sequential}. \\
\\
The strengths and limitations of each different adaptive design have been examined in various studies e.g. ALM is not as computational expensive as other sequential designs but tends to place many points in the boundaries of the design space \cite{beck2016sequential, gramacy2009, krause2008} whereas ALC, compared to ALM, is computationally more expensive but it performs better as it examines the effect of each point from the candidate set over the entire domain \cite{gramacy2009}. On the other hand, MICE outperforms other designs due to its computational efficiency and robustness \cite{beck2016sequential}. In the current study, both categories somehow are used: the initial input points and the input points that cover the entire search space, which are available for selection during the optimization process, are all sampled using LHD. To determine at which input point to evaluate the computer model next, MICE, which is described further in the next section as it is integrated into the optimization scheme, is used as a sampling criterion. 

\subsection{Bandit Setting}
The strategy followed to find the optimal value in the current work is to achieve a balance between exploration and exploitation. The idea is to gather more, or enough, information about the objective function by exploring the uncertain regions (high predictive variance) and then, make the best decision by exploiting, here optimizing, all the available information already known (high predictive mean). To achieve that, an acquisition function is used that not only controls the exploration-exploitation but also guides us on searching for the maximum \cite{brochu2010}. It is constructed based on the estimates obtained from the GP-based surrogate model \cite{Lizotte2008, chen2009}.\\
\\
The trade-off between exploration and exploitation has been studied in various studies within machine learning \cite{vcrepinvsek2013,ishii2002,kaelbling1996} and often seen as a multi-armed bandit problem \cite{auer2002,robbins1985,srinivas2010}. A multi-armed bandit problem is a sequential decision making problem where at each time step of a time horizon $T$, the algorithm chooses one of the available arms (i.e. candidate points) and calculates its reward \cite{robbins1985}. Under the GP optimization framework, the value of the unknown function at the chosen point $x_t$ is seen as the reward and the aim is to maximize the sum of rewards $\sum\nolimits_{t=1}^T f(x_t)$. To ensure that the strategy is performed well in a time horizon $T$, the loss (known as simple regret $r_t$) incurred at iteration $t$, when $x_t$ is chosen, is calculated at each time step as
\begin{align}
r_t = f(x^*) - f(x_t).
\end{align} 
Undoubtedly, a desirable asymptotic property of the optimization algorithm is to be no-regret
\begin{align}
\lim_{T\to\infty} \frac{R_T}{T} = 0,
\end{align} 
where $R_T$, known as cumulative regret, is the sum of the simple regrets after $T$ iterations. 

\maketitle
\section{optim-MICE: A Sequential-based Optimization Scheme}
An overview of the optimization scheme used and how the algorithm adaptively selects the input points at which the computer model is evaluated, in order to maximize the information gained over the input space, is given below.

\subsection{Confidence Region}
Under the GP framework, the predictive distribution at any input point $x_t$ is again a multivariate Gaussian distribution, $GP\left(\test{y}(x_t\right), \test{s}^2(x_t))$. A confidence region in which the unknown function $f$ is included with high probability is defined based on the well-known confidence interval. Using the Upper Confidence Bound (UCB), as an acquisition function, the confidence bounds are constructed as:
\begin{equation}
\begin{aligned}
\test{f}_t^{+}(x) &= \test{y}_{t}(x) + \sqrt{\beta_{t}} \test{s}_{t}(x)\\
\test{f}_t^{-}(x) &= \test{y}_{t}(x) - \sqrt{\beta_{t}} \test{s}_{t}(x),
\end{aligned}
\end{equation}
where $f_t^+$ is the upper bound, $f_t^-$ is the lower bound, $\test{y}(x_t)$ is the predictive mean and $\test{s}^2(x_t)$ the predictive standard deviation. The width of the confidence region is regulated by $\beta_t$ which also controls the trade-off between the exploration and exploitation.\\
The first input point, $x_t^1$, of each batch, is chosen based upon the UCB policy and it is the one that maximizes the upper bound,
\begin{equation}
x_t^1 = \underset{x \,\in \,\mathcal{X}}{\arg\max} \,\, \test{f}_t^+(x). 
\end{equation}

\begin{figure}
\centering
  \includegraphics[width=0.8\columnwidth]{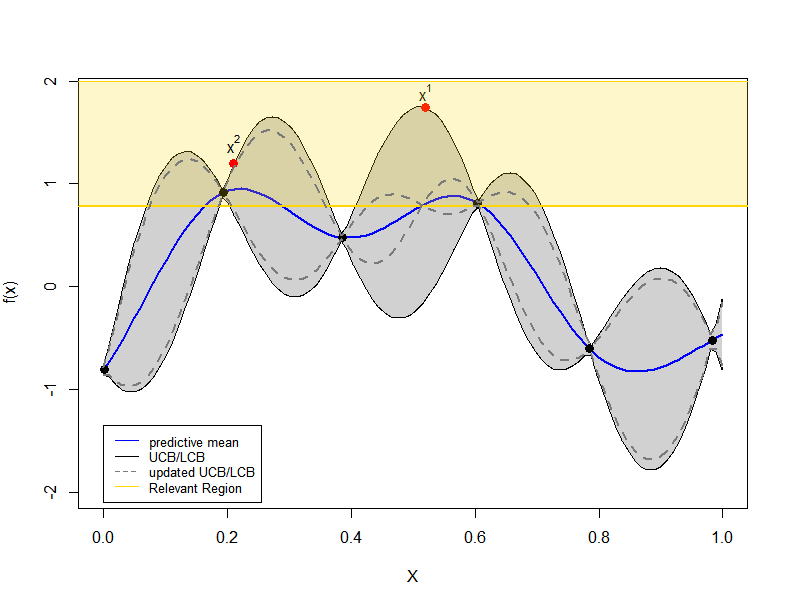}
\caption{Illustration of optim-MICE applied in $1$-dimensional test function. The grey area shows the confidence region and is bounded by $\test{f}_t^+$ and $\test{f}_t^-$. The first input point, $x^1$, is chosen based on the UCB-policy. The $\mathit{y}^{\bullet}_t$ is represented by the horizontal yellow line whereas the relevant region, $\mathfrak{R}_{t}$, is the yellow area. The black dashed lines shows the updated upper and lower bounds after having selected $x^1$. The second input point $x^2$ is chosen using the Pure Exploration strategy.}
\label{fig:1}       
\end{figure}

\subsection{Relevant Region}
A further reduction of the confidence region is obtained where the true optimum, $x^*$, of $f$ belongs with high probability. The relevant region, $\mathfrak{R}_{t}$, is defined as
\begin{equation}
\mathfrak{R}_{t} = \Big\{x \in \mathcal{X}\ | \ \test{f}_t^+(x) \geq \mathit{y}^{\bullet}_t \Big\}, 
\end{equation}\
where $\mathit{y}^{\bullet}_t$ is the lower confidence bound on the maximum,  $\mathit{y}^{\bullet}_t = \test{f}_{t}^-(x^{\bullet})$ and $x_t^{\bullet} = \underset{x \,\in \,\mathcal{X}}{\arg\max} \,\, \test{f}_t^-(x)$. At every iteration $t$, only the locations that might contain the true optimum of the unknown function $f$ with high probability are kept in the relevant region.

\subsection{Parallel Evaluations using MICE}
Applying the parallel strategy, we are able to choose a batch of $K$ input points at each iteration $t$. The $K-1$ remaining input points are selected via Pure Exploration. We restrict our attention to the relevant region $\mathfrak{R}_t$. The objective at this step is to maximize the information gain about the unknown function by selecting the most appropriate input points.\\
\\
In order to choose the next input point, we calculate the MICE criterion  \cite{beck2016sequential} for each point available for selection. The point that is selected and added in the batch is the one that maximizes the MICE criterion, namely the one that maximize the mutual information between the chosen design points and input points which have not yet been selected. MICE is based on the information theoretic mutual information criterion, given by Cover and Thomas \cite{Cover2006}, a measure of the information
contained in one random variable about another. For all $1 < k < K$, using a greedy strategy in which the new input points are selected one by one, 
\begin{equation}
x_{t}^k =  \underset{x \,\in \,\mathfrak{R}_t}{\arg\max} \,\, \test{s}_t^{2}(x) / \test{s}_{G \setminus (t \cup x)}^{2}(x; \tau_s^2),
\end{equation}
where $\test{s}_t^2$ is the predictive variance after the next input point is chosen and included in the batch and $\test{s}_{G \setminus (t \cup x)}^{2}(x; \tau_s^2)$ denotes the predictive variance of the points that have not been selected yet. The factor $\tau_s^2$ referred as the nugget parameter and is added in the correlation matrix $K$. The predictive variance does not depend on the unknown function evaluation but only on the actual location of the next input point $x_t^k$. After choosing the $K-1$ input points, the uncertainty about the unknown function is reduced and the guess about the upper bound in the next iteration is improved. The overall procedure is illustrated in Fig. \ref{fig:1}.

\maketitle
\section{optim-MICE: Computational Efficiency}
\begin{figure}[t!]
\minipage{0.5\textwidth}
\centering{\text{Branin 2D}}
\vspace*{-2.8mm}
  \includegraphics[width=\linewidth]{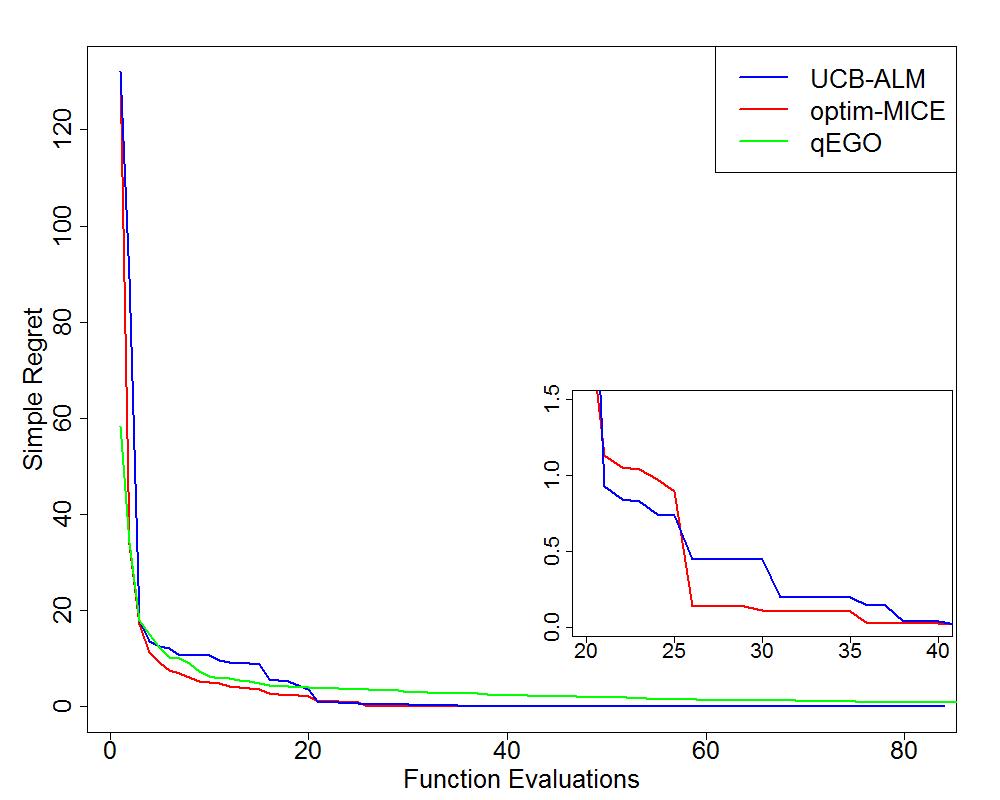}
    \vspace*{0.10cm}
\endminipage\hfill
\minipage{0.5\textwidth}%
\centering{\text{Rosenbrock 3D}}
\vspace*{-2.8mm}
  \includegraphics[width=\linewidth]{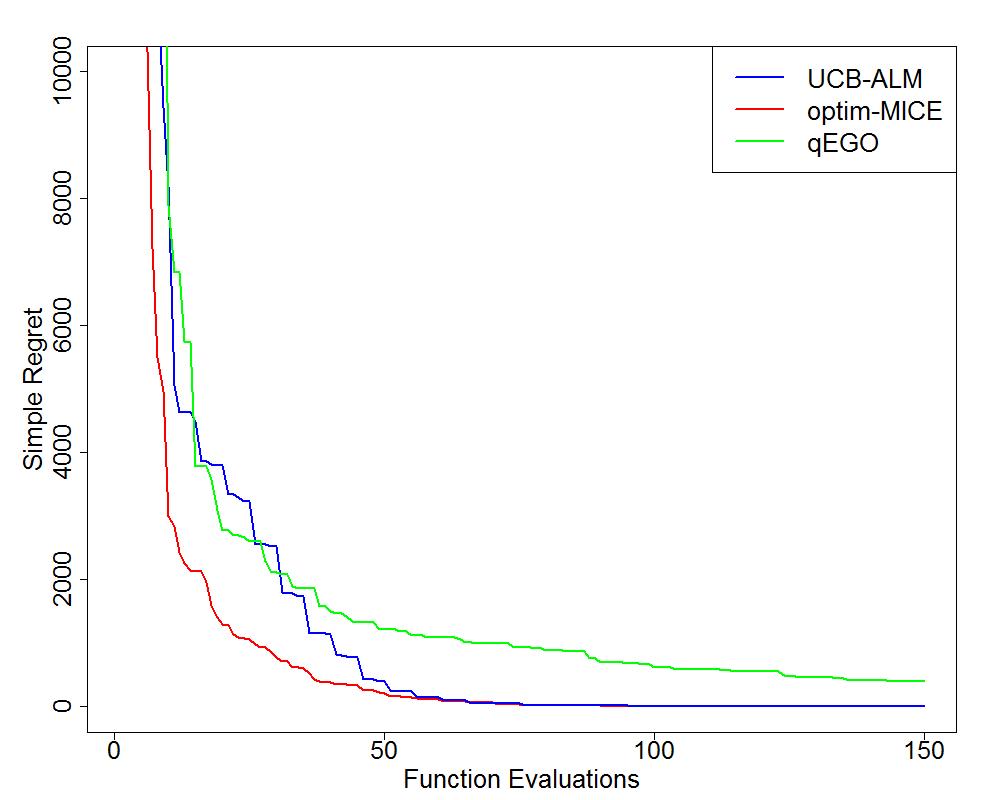}
  \vspace*{0.10cm}
\endminipage\\
\minipage{0.5\textwidth}
%\captionsetup{labelformat=empty}
  \includegraphics[width=\linewidth]{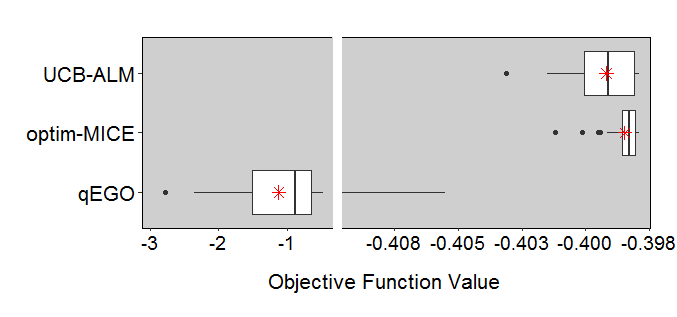}
      \vspace*{0.10cm}
\endminipage\hfill
\minipage{0.5\textwidth}
%\captionsetup{labelformat=empty}
  \includegraphics[width=\linewidth]{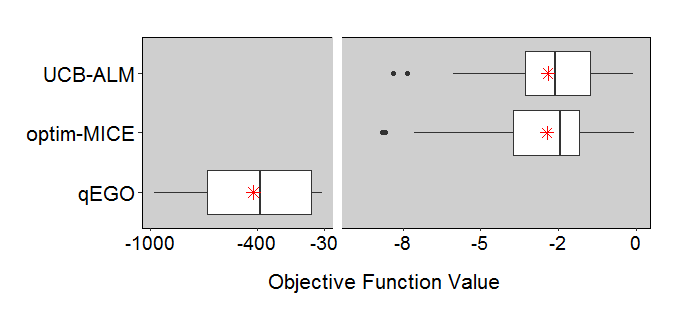}
      \vspace*{0.10cm}
\endminipage \hfill \\
\captionof{figure}{\textit{top}: Comparison of the mean simple regret with respect to the total function evaluations over 50 trials. \textit{bottom}: Summary of the best solution achieved with a gap in the rage of function values. Red stars show the mean best solution.}
\label{fig:2} 
\end{figure}

The empirical performance of optim-MICE is demonstrated in detail in \cite{mathikolonis2019surrogate}. The newly developed optimization scheme is compared with state-of-the-art heuristics such as GP-UCB-PE \cite{contal2013}, an algorithm which makes use of the UCB acquisition function and ALM as an experimental design, and the \textit{q}-points EI \cite{chevalier2013fast,ginsbourger2008multi}, which is based on the well-known Efficient Global Optimization (EGO) algorithm, on different optimization test functions. For convenience, the optimization schemes are referred to as UCB-ALM and qEQO, respectively. Specifically, the analysis held on 16 different computational experiments where the objective functions differ in physical properties, shapes and input dimensions. To keep consistency the algorithm settings in each optimization method are chosen to be the same whereas, to get an accurate measure of their performance 50 trials of each computational experiment are performed. Fig. \ref{fig:2} shows a comparison of the mean simple regret with respect to the total function evaluations performed during the optimization procedure for two test functions, the two-dimensional Branin function and the three-dimensional Rosenbrock function, and a summary of the best solution achieved in the 50 trials in box-plots.
Compared to the alternatives, optim-MICE needs less computational time to find the true optimum and achieve convergence of the regret. Regardless of the dimensionality and complexity of the objective function, the search process for identifying the uncertain region is more efficient and the regret decays faster compared to the other optimization algorithms \cite{mathikolonis2019surrogate}. The authors also discussed the effect of the main settings (number of iterations, batch size and the number of candidate points available for selection) on the overall performance of the algorithm. To get the most of optim-MICE, a balance between the number of iterations and the batch size is important. Choosing to perform just a few iterations but with a large batch size helps the search process at the initial stage and always ensures a good solution but the convergence of the regret can be slow \cite{mathikolonis2019surrogate}. In terms of the number of candidate points available for selection, things are more straightforward: the more points examined with the MICE criterion, the faster the maximum is obtained.

\maketitle
\section{Storm Surge Extrema: Results and Discussion}
BOSZ is a modular computer code, which in this study is treated as a black-box model. The computational expense of a Boussinesq-type model is relatively small in comparison to a full 3D model; however, thousands of individual runs can require substantial amounts of time even if the computation is executed only along a transect across the shoreline. The presented technique ensures that the variation of input variables efficiently influences the computer model's output in a way that the maximum bore or run-up heights are quickly achieved. The overall efficiency of the computer model is improved and a large amount of computational time is saved during the optimization procedure. To be more precise, the entire optimization task takes approximately 10 hours on total on a cluster of 4 cores whereas, each run of BOSZ takes about 2.5 minutes for our idealised set-up used here to demonstrate capabilities. For realistic geometries ideally modeled on a two-dimensional domain at high resolution each run could take even hours and the total computational time of performing such an optimization task would increase sharply.

\subsection{Variable Importance Assessment}
\begin{figure}[h!]
\centering 
  \includegraphics[width=1\linewidth]{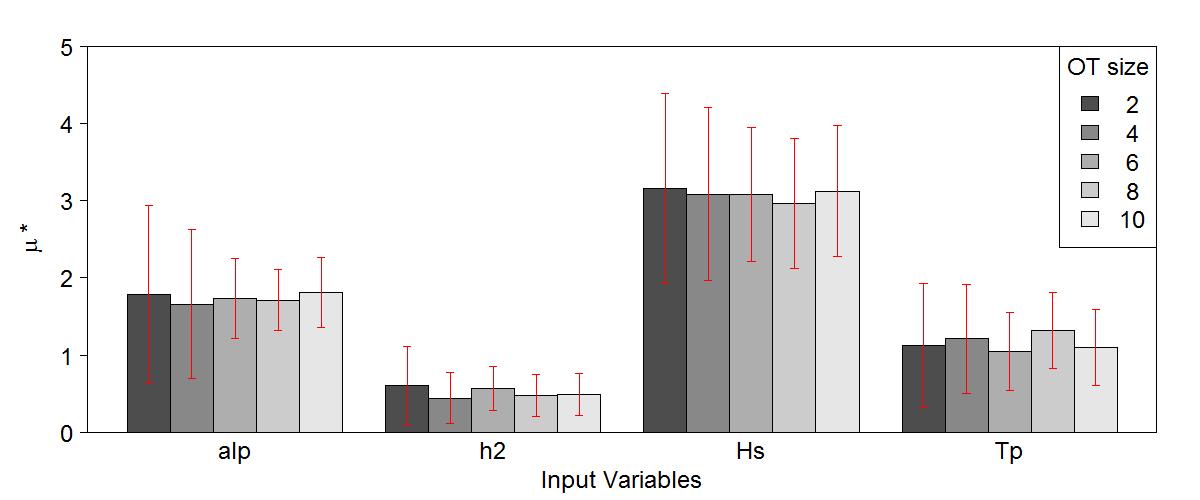}\\
  \includegraphics[width=1\linewidth]{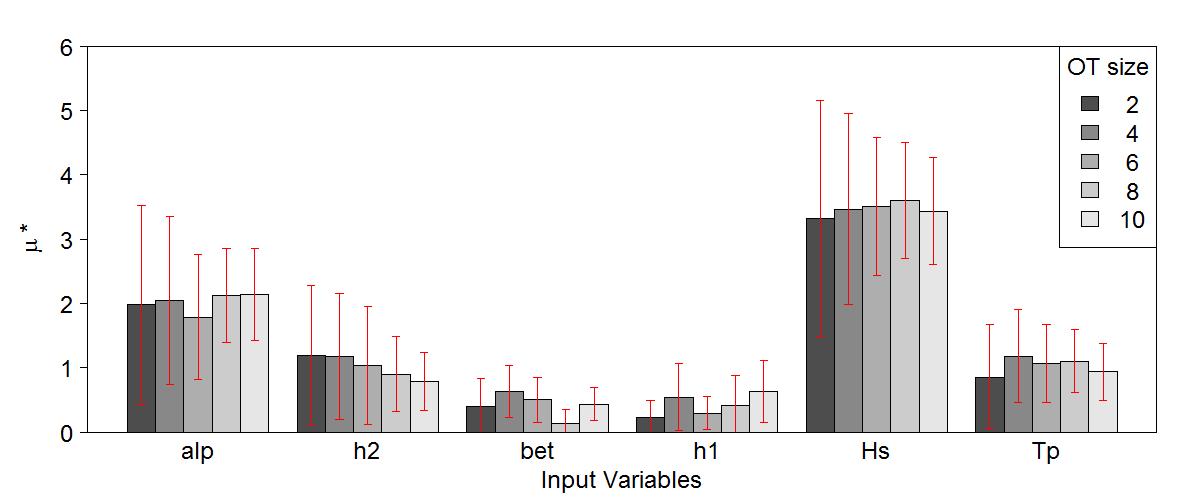}
  \caption{Robustness of the sensitivity results at different sizes of the optimal set of trajectories. Red lines show the error bars of the mean absolute elementary effects ($\mu^*$). \textit{Top panels}: for Bore height, \textit{Bottom panels}: for Wave run-up.}
\label{fig:3}
\end{figure}

\begin{figure}[h!]
\centering 
  \includegraphics[width=1\linewidth]{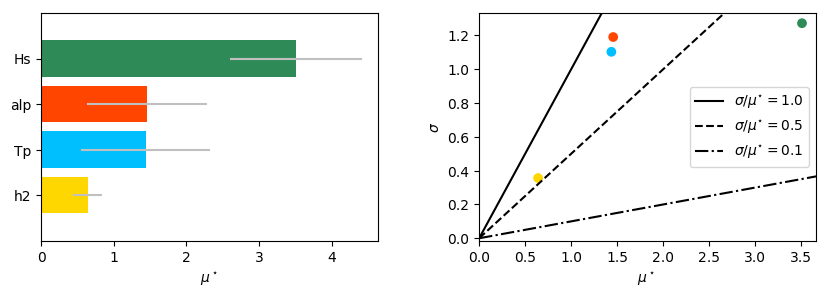}\\
  \includegraphics[width=1\linewidth]{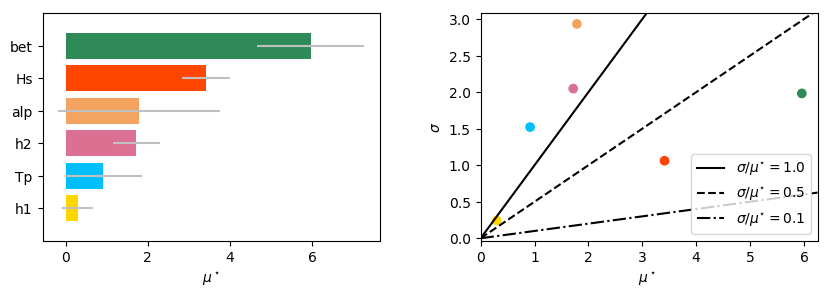}
 \caption{Results for the screening of input variables with the Morris method. Grey lines show the error bars of the mean absolute elementary effects ($\mu^*$). \textit{top}: for Bore height, \textit{bottom}: for Wave run-up.}
\label{fig:4}
\end{figure}

To find which input variables are significantly contributing to the model output, before the optimization starts, we choose to perform the Morris Method \cite{morris1991factorial}, an initial Sensitivity Analysis (SA), also known as a Screening technique, which is widely used in the area of computer experiments \cite{boukouvalas2014efficient,iooss2017efficient} and its effectiveness and efficiency has already been proved \cite{campolongo2007effective,sanchez2014application}.\\
\\
With a small number of model evaluations and without relying on strong assumptions about the computer model, the Morris method fast explores the model behaviour and ranks the input variables according to their impact on the model output. The input space for each input variable, as given in Tables \ref{tab:1} and \ref{tab:2}, is discretized in levels and then, a factorial sampling strategy is used to construct trajectories where one input variable is varied at a time by a pre-defined step $\Delta$, while keeping all others fixed. The sensitivity information is obtained by using the defined trajectories and computing for each input variable a number of incremental ratios, called Elementary Effect (EE). \\
\\
To be able to evaluate the global sensitivity of the model and obtain a global measure considering the whole input space, a set of \textit{r} different random trajectories needs to be constructed \cite{campolongo2011screening, sanchez2014application}. Having \textit{r} trajectories, a finite distribution of the EEs is formed which allows deriving statistical measures of the overall importance of each input variable: $\mu$, which assesses the influence of each input variable on the model output and measures the overall sensitivity and the standard deviation, $\sigma$, which measures the involved interactions and non-linearity effects of the $i$th input variable without allowing to make the distinction between the two cases, non-linearities and interactions. A challenging problem  is to choose the number of trajectories. Despite the fact that the analysis is more precise when $r$ is high, it increases the computational cost significantly. The literature suggests either a number for trajectories or different approaches on how to optimally choose $r$ however, the current work use the Optimal Trajectory (OT) approach: from a huge number of trajectories which are built at the beginning, we select the combination of $r$ trajectories with the highest "spread" in the experimental region using the Euclidean Distance (ED). More details can be found in \cite{campolongo2007effective}. \\
\\
Before starting the screening strategy, an experiment is performed with the aim to find an appropriate number of trajectories. Since the two optimization tasks differ in the number of input variables, the OT approach is applied in both cases. We compare the robustness of the sensitivity results while the size of the optimal set of trajectories is changing. To overcome the issue of the opposite signs, we use the mean value of the absolute EEs, which also indicates the importance of the input parameters \cite{campolongo2011screening}. The results are shown in Fig. \ref{fig:3}. Any value greater than $r=6$ for the bore height case and $r=8$ for the run-up case, would result in performing unnecessary model runs as the accuracy of the results is not improved.\\
\\
For completeness, the Morris screening strategy is applied in both cases. The input variables are ranked in order of importance and, according to the classification scheme proposed in \cite{sanchez2014application} which is based on the ratio $\sigma/\mu^*$, are characterised in terms of linearity ($\sigma/\mu^*<0.1$), monotony ($0.5>\sigma/\mu^*>0.1$) or possible interactions ($\sigma/\mu^*>1$). Fig \ref{fig:4} shows the Morris screening results for both cases.\\ 
\\
For the bore height case, all the input variables appear with significance influence, with the most influential one being the the significant wave height (\textit{Hs}). None of them show a strong linear effect or possible interactions with at least one other variable. A linear relationship between the water depth over the reef (\textit{h2}) and the bore height can be stated due to the relatively low standard deviation. For the wave run-up case, the most influential input variables are the beach slope (\textit{bet}) and the significant wave height (\textit{Hs}). The fore-reef slope (\textit{alp}), water depth over the reef (\textit{h2}) and peak period (\textit{Tp}) are less important but still with a significant influence on the model's output. On the other hand, the water depth offshore (\textit{h1}) can be classified as negligible, without any impact on BOSZ and therefore, can be fixed during the optimization procedure. Except of the beach slope and the significant wave height, where their behaviour can be characterized closer to linear, all the other input variables show a non-linear influence and/or interactions with other parameters ($\sigma/\mu^*>0.5$). \ 
\\

\subsection{Extreme Bore Height}
\begin{figure}[h!]
\centering
\includegraphics[width=.4830112\textwidth]{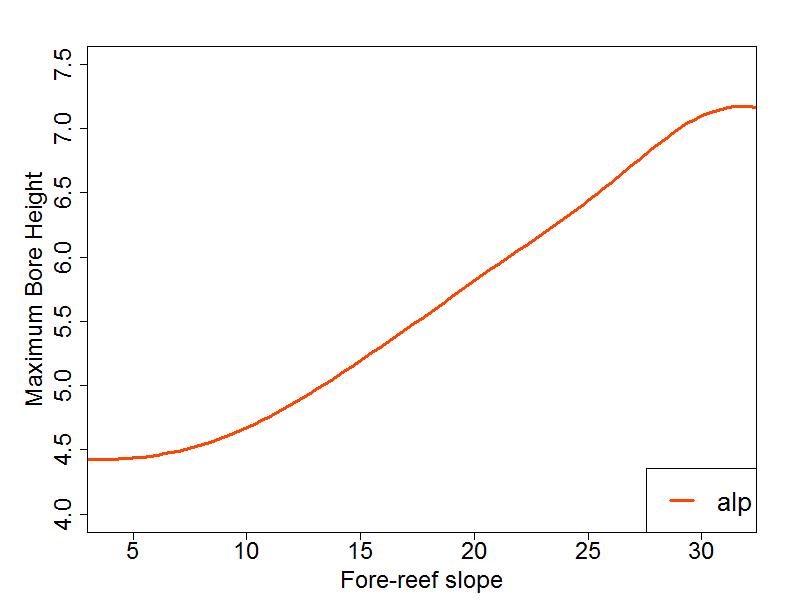}\quad
\includegraphics[width=.4830112\textwidth]{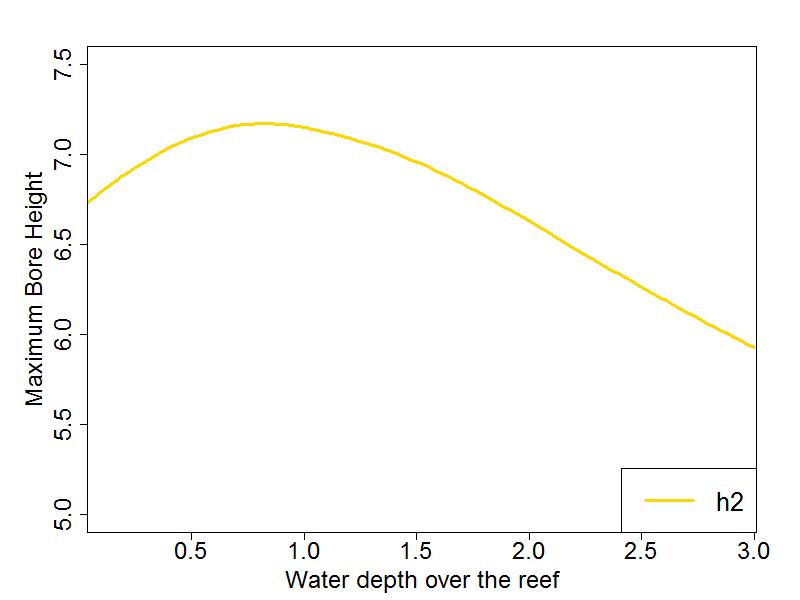}
\medskip
\includegraphics[width=.4830112\textwidth]{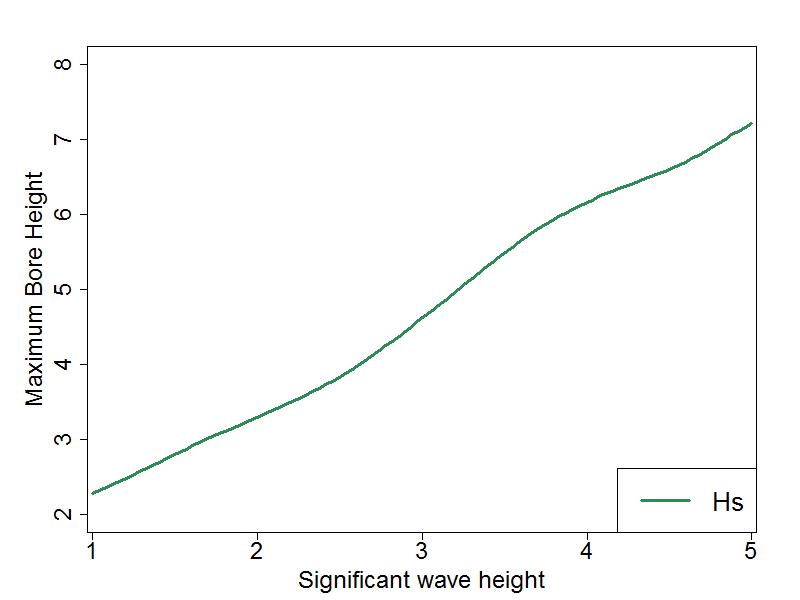}\quad
\includegraphics[width=.48300112\textwidth]{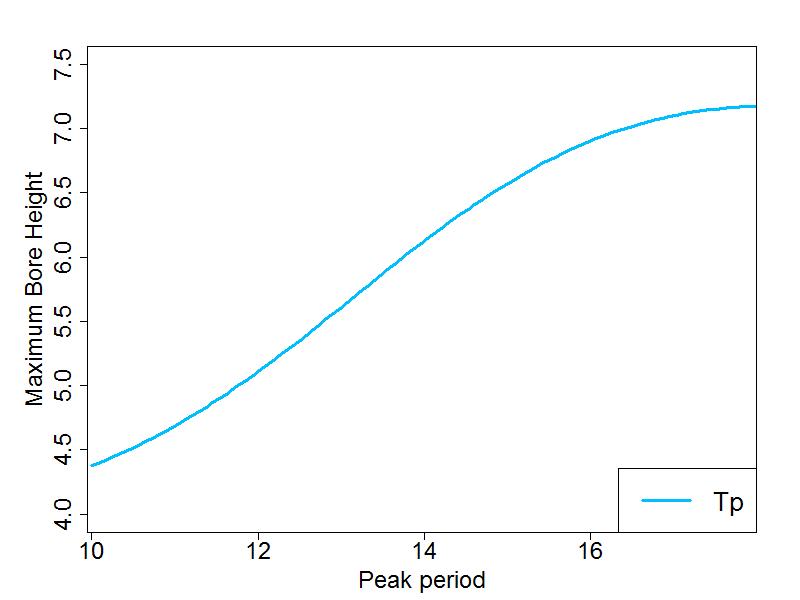}
\caption{Local sensitivity of the maximum bore height on the fore-reef slope ($alp$), water depth over the reef ($h2$), significant wave height ($Hs$), peak period ($Tp$).}
\label{fig:5}
\end{figure}

After performing 200 simulations and using optim-MICE, the maximum bore height obtained is 7m (Fig. \ref{fig:6}). To get a better understanding which of the input variables control the bore height and how, a local sensitivity analysis around the maximum bore height is performed using 500 runs of the last fitted Gaussian Process emulator. This allows us to carry out a near instantaneous sensitivity analysis (in less than 1 s), whereas running BOSZ around the maximum for 500 of inputs would be computationally costly, taking hours or days. The four panels in Fig. \ref{fig:5} shows the behavior of the maximum bore height over the reef flat as a function of the four input parameters reef slope (\textit{alp}), water depth over the reef (\textit{h2}), significant wave height (\textit{Hs}) and peak period (\textit{Tp}), under the condition that only one of the four parameters is changed and all others stay constant. \\
\\
It is not very surprising that the bore height increases with the significant wave height, i.e. the incident offshore wave energy. For a given set of peak period, reef slope and reef depth, the increase in bore height is close to linear with the value of significant wave height. However, the other variables involved show a different trend. So does the effect of an increasing peak period tend to taper off beyond $\text{Tp}\sim~16$ sec. The offshore wavelength increases quadratic with the peak period. As the peak period increases, the overall wave spectrum then contains significantly more energy from longer waves in comparison to a spectrum with a shorter peak period. At the same time, the overall composition of wave energy in the spectrum shifts towards longer wave groups. Such wave groups can be seen as reoccurring pulses (above mean water level) and lolls (below mean water level) of several waves due to the interaction of individual waves with each other. These wave groups are of much longer period than the individual swell waves and they are therefore called infra-gravity (IG) waves. IG waves can be of very different nature. A way to distinguish them is to classify them into two groups of bound and free IG waves. The IG waves resulting from the composition of the spectrum are inherently connected to the shape of the swell spectrum and they are therefore bound IG waves. In contrast, IG waves released after wave breaking are mostly free waves. In many cases, the presence of large wave groups also results in large breaking waves and run-up. In case of the maximum bore height it is very likely that the highest values are reached when the most energetic wave group is present. For the tested configurations, the increase in wave groups does not lead to a steady increase in observed bore height after wave breaking occurs over the reef flat.\\
\\
A similar, but yet more drastic effect can be observed when only the water depth over the reef is changed. Very shallow water over the reef does not lead to the largest bore heights, mainly because the height of a hydraulic jump depends on the water level at both sides of the discontinuity and the height of the jump itself. Over nearly dry bed, waves move as sheet flow rather than as a pronounced bore. However, for our particular configuration, water depths larger than $\sim ~1m$ have an adverse effect on the bore height. This is due to the fact that the incoming energy is distributed over a larger mass of water. Instead of having a jamming effect, the energy is absorbed by a larger mass of water that reduces the height of the bore.\\
\\
The influence of the reef slope on the bore height follows a different trend in comparison to the other parameters. Starting off with a very gentle slope (right side of panel 1 in Fig 5), the zone where waves break is relatively wide. The water depth is a controlling factor for the initiation of wave breaking, i.e. individual waves of various heights start breaking in different water depths. With a gentle bathymetry slope, this leads to a wide spread area where wave breaking is possible and, in turn, a more pronounced dissipation zone. As the slope increases, the horizontal area of wave breaking shrinks and most waves tend to break around the same position over the slope. At the same time, wave shoaling starts to become effective and increases the wave height before breaking. Shoaling is a group wave process, i.e. it acts on the underlying bound IG waves. With very steep slopes, parts of the wave energy is reflected and the bore height behind the breaking zone does not further increase. 

\subsection{Extreme Wave Run-Up}
\begin{figure}[h!]
\minipage{0.5\textwidth}
\centering{\text{Bore Height}}
\vspace*{-2.8mm}
  \includegraphics[width=\linewidth]{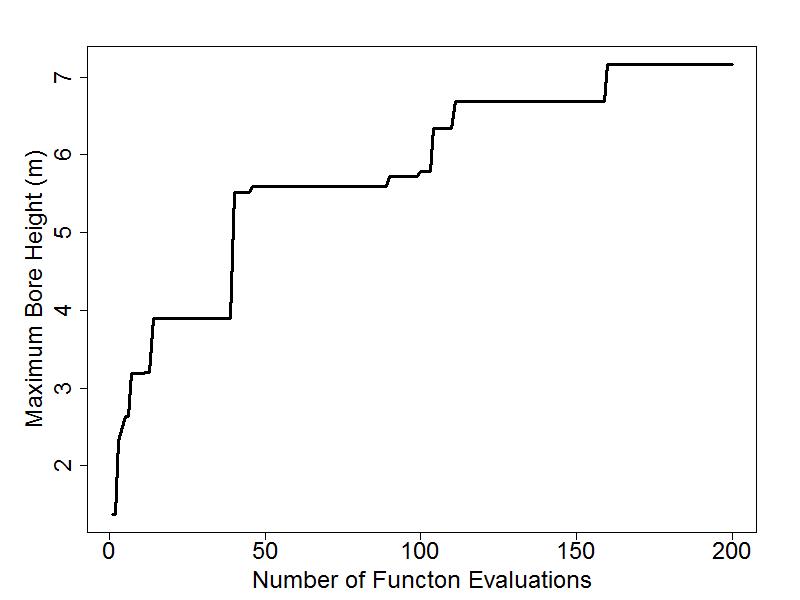}
    \vspace*{0.10cm}
\endminipage\hfill
\minipage{0.5\textwidth}%
\centering{\text{Wave Run-up}}
\vspace*{-2.8mm}
  \includegraphics[width=\linewidth]{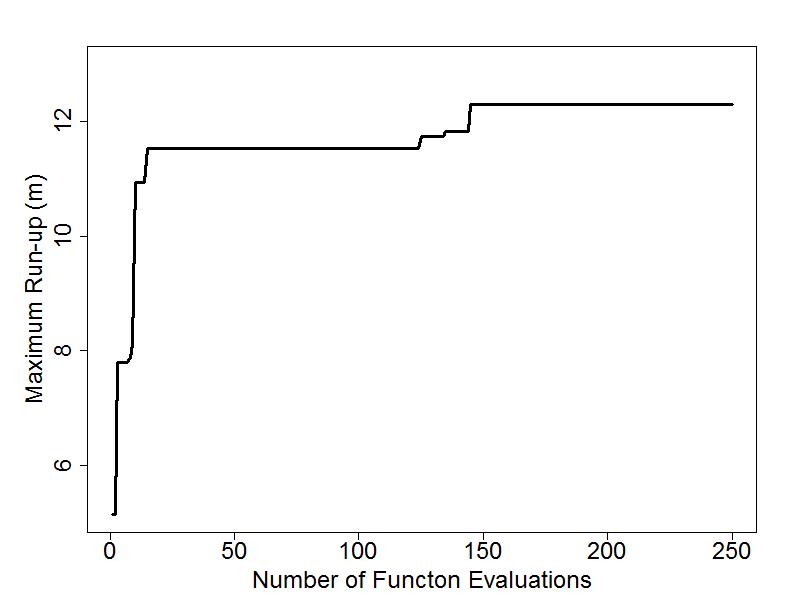}
  \vspace*{0.10cm}
\endminipage\hfill \\
\captionof{figure}{Optimization of BOSZ using the optim-MICE algorithm. \textit{left}: Maximum Bore Height versus total function evaluations. \textit{right}: Maximum Run-up versus total function evaluations.}
\label{fig:6} 
\end{figure}

\begin{figure}[ht]
\centering
\includegraphics[width=.4830112\textwidth]{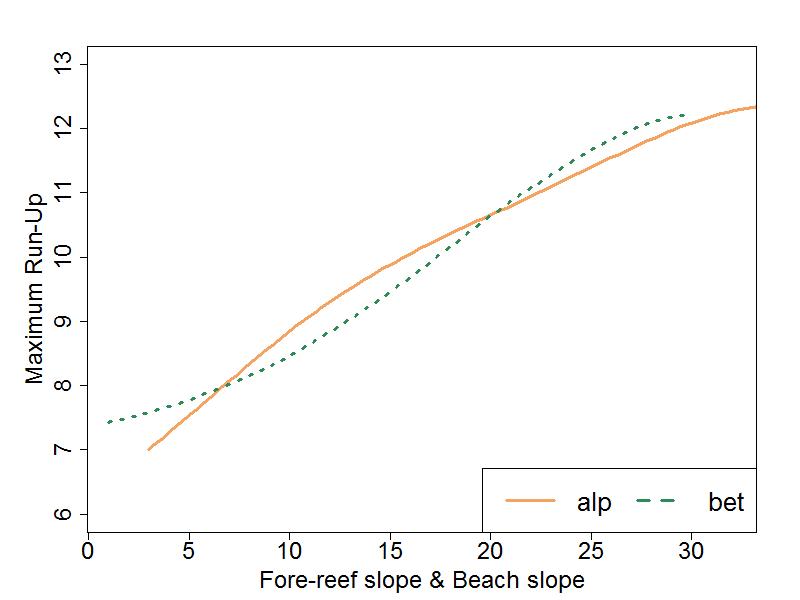}\quad
\includegraphics[width=.4830112\textwidth]{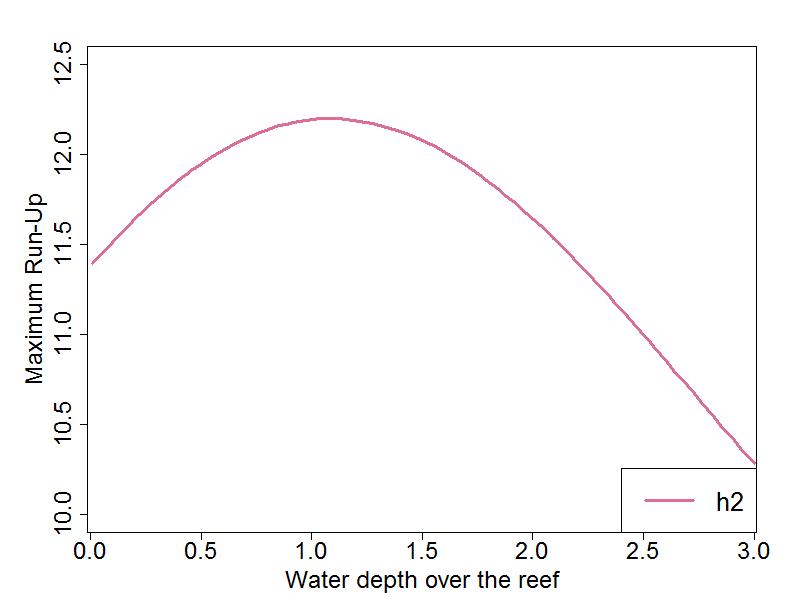}
\medskip
\includegraphics[width=.4830112\textwidth]{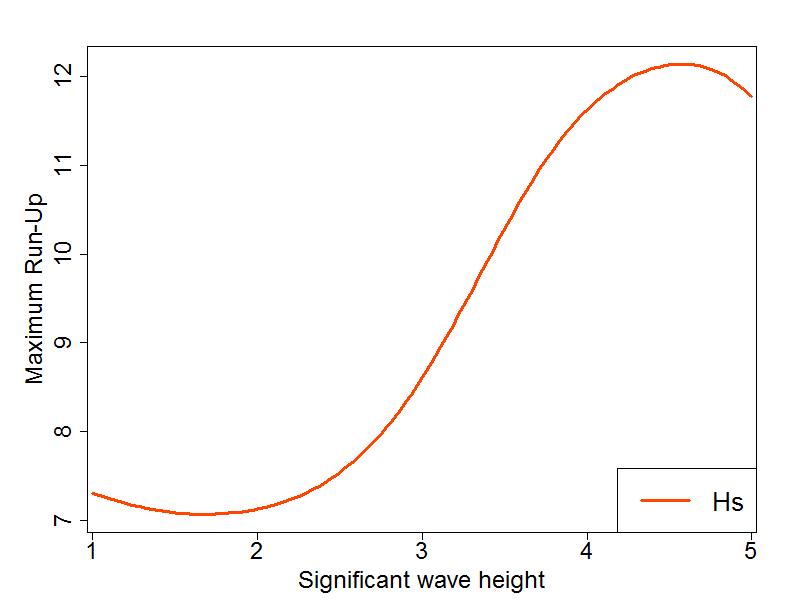}\quad
\includegraphics[width=.48300112\textwidth]{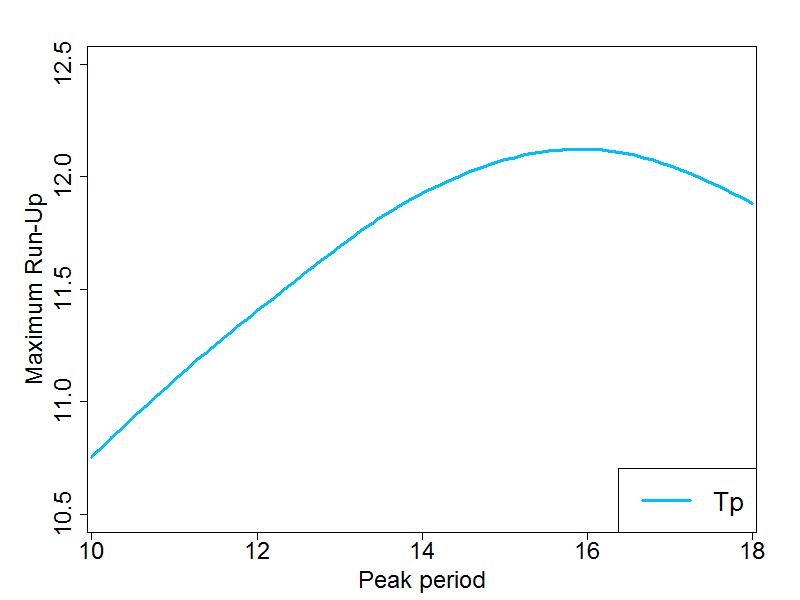}
\caption{Local sensitivity of the maximum run-up on the fore-reef slope ($alp$), beach slope ($bet$), water depth over the reef ($h2$), significant wave height ($Hs$), peak period ($Tp$).}
\label{fig:7}
\end{figure}

After performing 250 simulations and using optim-MICE, the maximum wave run-up obtained is 12m (Fig. \ref{fig:6}). The behavior of the maximum run-up in dependence of the remaining parameters can be analyzed in a similar fashion as in the bore height case. Using the last Gaussian process emulator fitted, a local sensitivity analysis around the maximum run-up is performed. The model was setup in an almost identical way compared to the bore height analysis - with the exception that a dry slope was added to the right side of the reef slope where waves can freely run up on after having propagated over the reef flat. The four panels in Fig. \ref{fig:7} show the behavior of the maximum run-up over the dry slope as a function of the other four input parameters significant wave height ($Hs$), peak period ($Tp$), water depth over the reef ($h2$), and bathymetry slopes, fore-reef slope ($alp$) and beach slope ($bet$), under the condition that only one of the four parameters is changed and all others remain constant. \\
\\
In contrast to the maximum bore height, the highest run-up does not exhibit a near-linear relationship with the incoming wave height ($Hs$). The maximum and minimum run-up cannot be found for the highest and lowest significant wave height of $5m$ and $1m$ respectively, but instead they occur at about 4.5m and 1.7m. In between the two values, a strong increase in run-up can be observed. \\
\\
The dependence of the run-up on the peak period ($Tp$) and on the water depth over the reef ($h2$) follows a similar trend as what can be seen for the maximum bore height. Especially for the water depth of the reef, the highest run-up coincides approximately with the presence of the highest bore over the reef. This is somehow intuitive because there is no obstruction of the propagating bores on their way towards the run-up slope; i.e. an approaching large bore will most likely run up far on the slope unless an opposing flow from a previous drawdown stage reduces the flow momentum.\\ 
\\
For the two slopes (fore-reef and run-up slope) we can conclude that a gentle slope generally leads to smaller run-up values than a steep slope. The reasons are different from what influences the maximum bore height. The lower run-up heights over gentle topographic slopes result mainly from the fact that wave breaking can take place over a longer duration and distance. Though, no bottom friction was implemented in the study, the effects of frictional loss are more pronounced over gentle slopes than over steep slopes. In other words, if frictional losses were accounted for, the dashed line in the upper left panel of Fig. 9 would be steeper. 

\subsection{Total effect of dependencies on Run-up}
Another way to look at the influence of the maximum wave run-up on a dry slope is the non-dimensional surf similarity parameter or Iribarren number expressed as $ Ir = \tan alp/\sqrt{Hs/Lo}$, where $L_0 = 2Tp^2/2\pi$. Originally designed to give an estimate of wave breaker type and wave breaking intensity as a function of beach slope and offshore wave steepness. In general, low values indicate a gentle ''spilling'' type of wave breaking, whereas high Iribarren numbers show evidence of plunging and even collapsing breakers - a type of wave breaking that occurs instantaneously in contrast to the long-lasting spilling breakers. For a fixed slope, a lower wave steepness results in more abrupt wave breaking types. An extreme example is a tsunami wave, which usually is very long but of low amplitude. Once it approaches shallow water, it shoals over a long distance and ultimately breaks abruptly into a bore.\\
\\
With respect to the presented case study, we generally observe high run-up values for low Iribarren numbers. Fig. 8a shows that the maximum run-up occurs for rather low ratios of reef depth to offshore wave height (significant wave height $Hs$ is high in comparison to the water depth over the reef $h2$) in combination with low Iribarren numbers. A similar behavior can be observed in Fig. 8b where low Iribarren numbers lead to high run-up values if the ratio of bore height over the reef to offshore wave height is small. Both trends can be explained through the mechanisms of energy dissipation in breaking waves. A low bore height to significant wave height indicates rather low dissipation rates of breaking waves in relation to the initial energy level. As the bores approach the run-up slope, they describe spilling rather than collapsing breakers and the energy dissipation is not instantaneous. The same is true for low ratios of water depth over the reef to offshore wave height (Fig. 8a). This leads to relatively high run-up heights.\\
\\
Along the same lines, Fig. 8c shows that high run-up values can be found for low Iribarren numbers and low ratios of fore-reef slope to significant wave heights. In case of large incoming waves over gentle slopes, the waves approach rather as spilling breakers over a broad surf zone. It should be pointed out that the run-up values are distributed over a wide range of surf similarity values, i.e. low Iribarren numbers do not automatically lead to high run-up values. However, the largest run-up values fall into the category of low Iribarren numbers and no scenario in this case study shows very high run-up heights in combination with high surf similarity parameters.

\begin{figure}[h]
\minipage{0.5\textwidth}
%\captionsetup{labelformat=empty}
%\caption{}
\includegraphics[width=\linewidth]{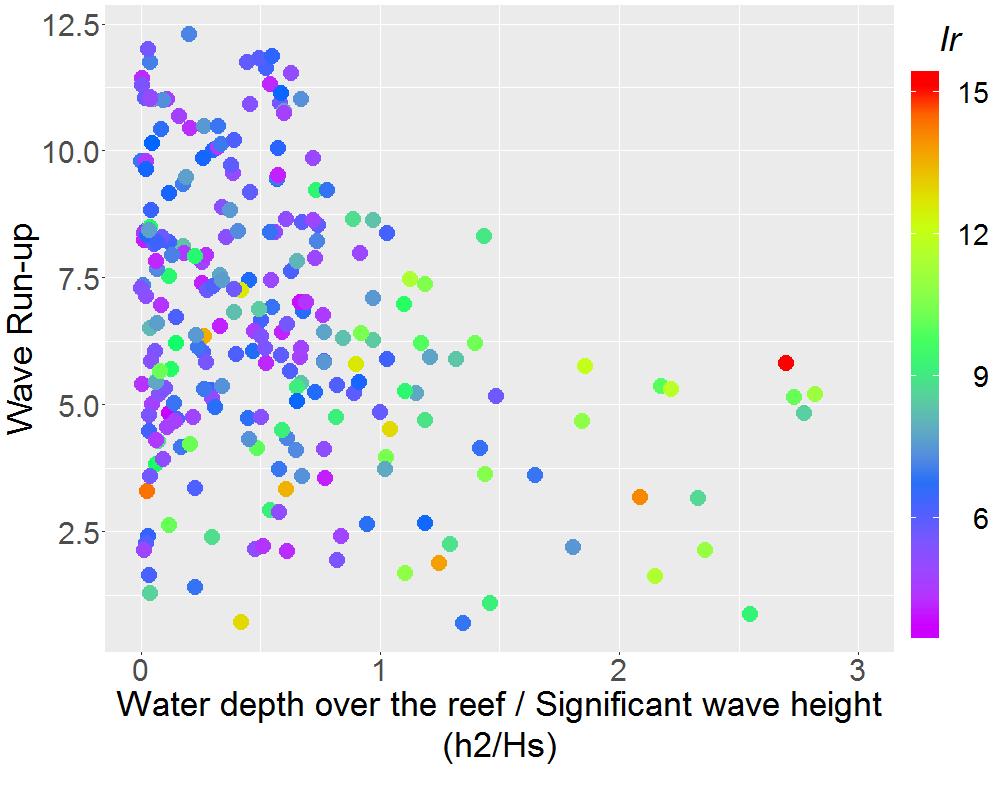}
  \centering\footnotesize{(a)}
  \vspace{0.50cm}
\endminipage\hfill
\minipage{0.5\textwidth}%
%\captionsetup{labelformat=empty}
%\caption{}
  \includegraphics[width=\linewidth]{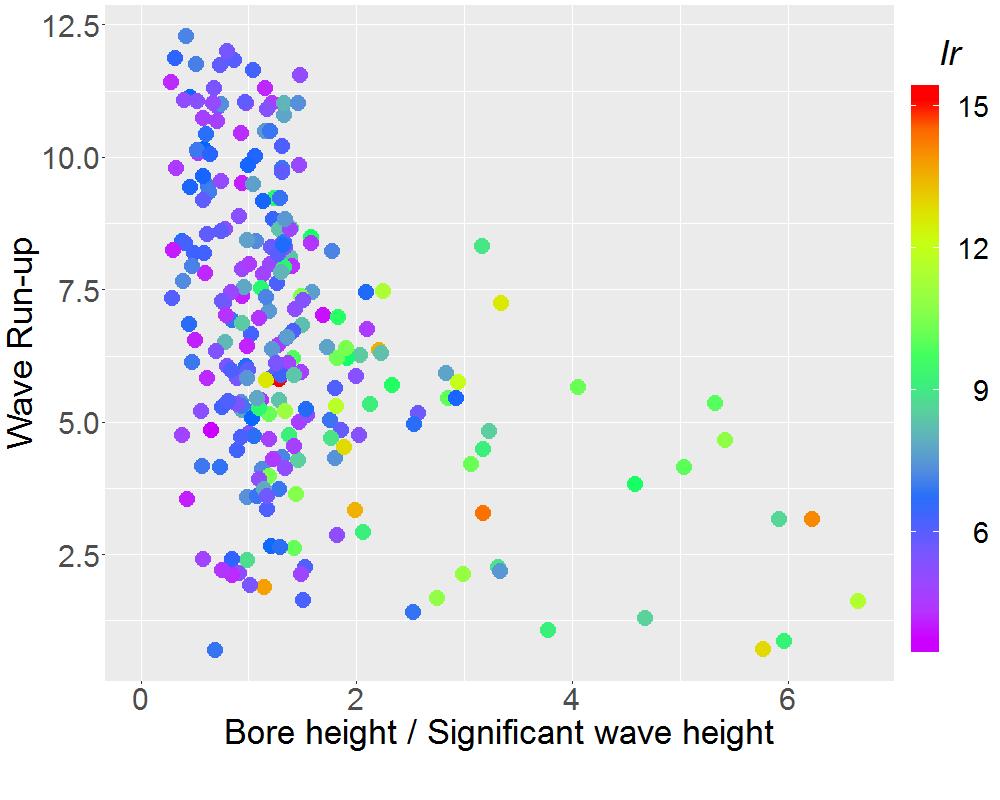}
  \centering\footnotesize{(b)}
  \vspace{0.50cm}
\endminipage \hfill\\
\centering
\minipage{0.5\textwidth}%
%\captionsetup{labelformat=empty}
%\caption{}
  \includegraphics[width=\linewidth]{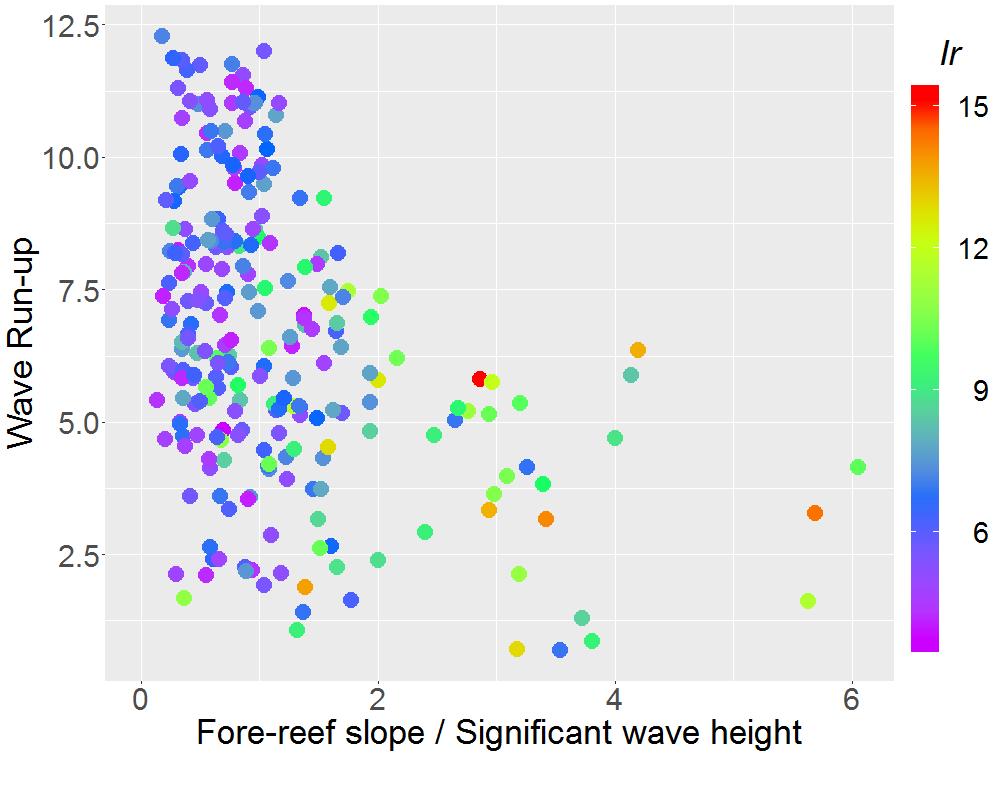}
  \centering\footnotesize{(c)}
  \vspace{0.50cm}
\endminipage \hfill \\
\captionof{figure}{Wave run-up as a function of (a) water depth over the reef to significant wave height, (b) bore height to significant wave height and (c) fore reef slope to significant wave height. The bar indicates the Iribarren number \textit{Ir} (the surf similarity parameter) calculated with the fore-reef slope, significant wave height and wavelength.}
\label{fig:8} 
\end{figure}

\maketitle
\section{Conclusion}
In this paper we identify the extrema of coastal storm waves over a reef-type bathymetry, for two quantities: breaking wave height and local wave run-up. The hypothetical setting is similar to what governed the wave event that destroyed the town of Hernani during Typhoon Haiyan. The capability of the BOSZ model in handling various wave processes, gave us the opportunity to explore different aspects of storm wave extrema, such as the maximum bore height and the maximum run-up, at a local level. Due to the computational complexity of the model, finding the combination of the controlling parameters that leads to the worst-case scenario requires a large number of individuals runs. To find the combination of input variables that creates possible large storm wave run-ups with the minimum computational effort, we employed optim-MICE, a newly developed optimization scheme which is based on the Gaussian Process and an information theoretic mutual information measure. Specifically, optim-MICE explores the entire input space and focuses on the region where the maximum belongs in with high probability. The simulator is only evaluated at a batch of input points that are chosen to yield the maximum information about it.\\
\\
Taking advantage of the computational efficiency of the optim-MICE algorithm, the maximum run-up and bore height are obtained in the lowest possible number of function evaluations (model runs). Using optim-MICE, we efficiently identified and explored the important regions faster compared to the alternative optimization schemes. This was also demonstrated in different experiments performed regardless of dimensionality and complexity of the objective function (Fig. \ref{fig:2}). Overall, the computational analysis performed in \cite{mathikolonis2019surrogate} shows that the total individual runs needed to find the maximum, or a solution close to the maximum, is increased by $20\%$ for GP-UCB-PE, and $75\%$ for qEGO, when compared to optim-MICE. Considering now only the cases with higher dimensions, on average, getting a solution close to the true optimum with GP-UCB-PE was required to perform 50 more function evaluations and 140 with qEGO, compared to optim-MICE (with a 250 run budget). Translating this into computational time, and knowing that the current study needed about 10 hours on a cluster of 4 cores, such an optimization task would be expected to be completed in approximately 12 and 16 hours if GP-UCB-PE and qEGO would have been used, respectively. The more complex the computer set-up, e.g. for realistic coastal hazard assessments with one run taking hours not minutes, the more computational resources are required to complete the overall optimization task and therefore, it is crucial to ensure that a good solution can be achieved in the lowest possible computational time. Also, as shown in \cite{mathikolonis2019surrogate}, within a predefined number of function evaluations, the overall performance of optim-MICE compared to the alternatives is better at imparting confidence that the maximum, or a solution close to the maximum, can be achieved.\\
\\
The current study focuses on storm waves to better understand the factors that contribute to nearshore breaking wave heights and local run-up on the beach. The optim-MICE algorithm could be extended to a multi-objective optimization context or to other types of waves, e.g. tsunami waves. It could for instance be used to optimize even more complex simulations for multiple outputs at once, say for coastal wave heights and currents across an entire region, not one output type at one location at a time as in this paper. \\
\\
This case study uses an idealized bathymetry whose main features characterize a fringing reef. The computation of a suite of typical storm waves over this bathymetry shows that maximum breaking wave heights and runup on a straight beach behind the reef follow particular trends. Most important, the run-up on the beach after the wave breaking process is not linearly related to the incoming wave energy, whereas the breaking wave height is close to linearly related to the height of the approaching waves. The water depth over the reef is a strong controlling parameter for both breaking wave height and run-up. Counter-intuitively, the highest values do not occur at the highest water level, but in fact at relatively low levels (around 1m in our case study). This can be explained with the generation of infragravity waves through wave breaking and their effect on the nearshore dynamics, which are highly nonlinear processes.\\
\\
Finally, With the presented optimization algorithm, it is possible shorten the computation time significantly that is usually required to run through a permutation of scenarios in order to find a worst case scenario. This approach will greatly help assess quickly the potential of coastal hazards and thus improve future hazard mitigation efforts.

\end{document}